\documentclass[aps,prl,twocolumn,showpacs,floatfix,superscriptaddress]{revtex4}
\usepackage{graphicx}
\usepackage{bm}
\usepackage{amsmath}
\begin{document}
\title{Nonequilibrium Dephasing in an Electronic Mach-Zehnder Interferometer}
\author{Seok-Chan Youn}
\affiliation{Department of Physics, Korea Advanced Institute of
Science and Technology, Daejeon 305-701, Korea}
\author{Hyun-Woo Lee}
\affiliation{PCTP and Department of Physics, Pohang University of
Science and Technology, Pohang, Kyungbuk 790-784, Korea}
\author{H.-S. Sim} \email{hssim@kaist.ac.kr}
\affiliation{Department of Physics, Korea Advanced Institute of
Science and Technology, Daejeon 305-701, Korea}
\date{\today}
\begin{abstract}

We study nonequilibrium dephasing in an electronic Mach-Zehnder
interferometer.
We demonstrate that
the shot noise at the beam splitter of the interferometer
generates an ensemble of
nonequilibrium electron density configurations and that electron
interactions induce configuration-specific phase shifts of an interfering
electron.
The resulting dephasing exhibits two characteristic
features, a lobe pattern in the visibility and phase jumps of $\pi$,
in good agreement with experimental data.

\end{abstract}


\maketitle

{\it Introduction.} --- An electronic analog of optical Mach-Zehnder
interferometry (E-MZI) has been recently realized
\cite{Ji,Neder_EXP,Litvin,Roulleau} by utilizing edge channels of
integer quantum Hall (IQH) liquids. As it is one of elementary types
of interferometry, it can serve as an important probe of electronic
coherence \cite{Roulleau2} and entanglement
\cite{Samuelsson,Sim,Neder_2ptl}.

The E-MZI has a simple setup consisting of two arms and two beam
splitters. Recent experiments \cite{Neder_EXP,Roulleau} on it,
nevertheless, revealed puzzling behavior that is hard to understand
within a noninteracting-electron description \cite{Chung}; the
interference visibility of the differential conductance shows
bias-dependent lobe patterns, accompanied by phase jumps of $\pi$ at
the minima of the lobes. There may exist some unnoticed fundamental
physics behind it.

Electron-electron interactions may be important for the puzzling
nonequilibrium behavior. Interaction effects were studied
\cite{Sukhorukov,Chalker} in the tunneling regime by using
bosonization methods. In Ref. \cite{Sukhorukov}, interactions
between an E-MZI channel and an additional one outside the E-MZI
were considered and the resulting resonant plasmon excitations were
proposed as an origin of the puzzle. On the other hand, roles of the
shot noise of an additional detecting channel were addressed
\cite{Neder_EXP2,Neder_TH} to understand similar lobes found in a
related experiment \cite{Neder_EXP2}.



\begin{figure}[b]
\includegraphics[width=0.45\textwidth,height=0.22\textheight]{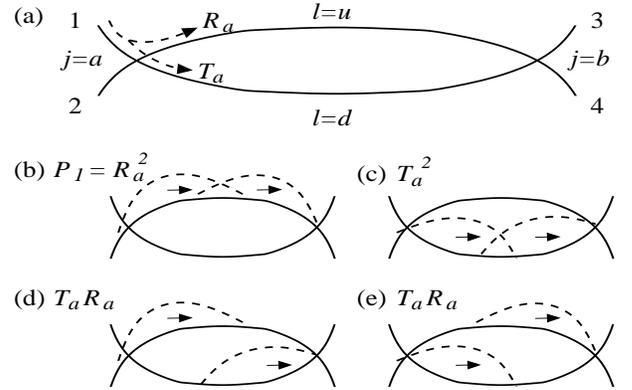}
\caption{(a) Schematic E-MZI setup. It has two beam splitters
$j=a,b$ with transmission probabilities $T_j$ and $R_j$, two arms $l
= u,d$, two sources $i=1,2$, and two drains $i=3,4$. Only the source
$1$ is biased. (b-e) Schematic nonequilibrium ensemble of electron
density configurations in the two arms resulting from the shot noise
at the splitter $a$ when the two arms have two nonequilibrium
electrons. The configurations can be described by two packets
(dashed lines) propagating towards the drains as marked by solid
arrows. The probability $P_m$ of each ensemble element $m$ is
marked.} \label{E-MZIsetup}
\end{figure}

In this work, we propose an intrinsic mechanism for the puzzling
behavior, which does not require additional channels outside the
E-MZI. A key observation is that the shot noise at the input beam
splitter of the E-MZI generates an {\em ensemble of nonequilibrium
electron density configurations} in the two arms. Then the electron
interaction {\em within each arm} induces configuration-specific
phase shifts of an interfering electron, and the ensemble average of
the phase shifts leads to
nonequilibrium dephasing. The combined effect of the shot noise and
the interaction results in lobe patterns and phase jumps, which
agree with experimental data \cite{Neder_EXP,Roulleau}. We use a
{\em wave-packet} picture to describe the nonequilibrium density and
treat the interaction phenomenologically at zero temperature. The
inter-arm interaction is ignored.

{\it E-MZI setup} --- The E-MZI consists of two sources $i = 1,2$,
two drains $i=3,4$, two beam splitters $j = a, b$, and two arms $l =
u,d$ of length $L_l$ [Fig. \ref{E-MZIsetup}(a)]. The source $i=1$ is
biased by $|e|V$, while $i=2$ is unbiased. Transmission
probabilities at the splitter $j$ are $T_j$ and $R_j (= 1-T_j)$.
Each arm consists of a single edge channel of the IQH liquid with
the linear energy dispersion characterized by constant group
velocity $v_F$. We first consider the simple case with $L_u=L_d
\equiv L$, and later discuss the case with $L_u \neq L_d$. There are
two time scales, electron flight time $t_{\textrm{fl}} \equiv L /
v_F$ through an arm and time separation $\tau_V = 2 \pi \hbar /
(|e|V)$ between successive injections of {\em nonequilibrium}
electrons (with single-particle energy $\in [0, |e|V]$) from the
source $1$. Their ratio gives the average number of the
nonequilibrium electrons in the two arms at a given time,
\begin{eqnarray}
N = \frac{t_{\textrm{fl}}}{\tau_V} = \frac{L |e| V}{2 \pi \hbar
v_F}. \label{AVE_N}
\end{eqnarray}

{\it Nonequilibrium density configurations.} --- At zero
temperature, the nonequilibrium state of the E-MZI can have the two
equivalent forms in the noninteracting limit,
\begin{equation}
|\Psi (t) \rangle_{\rm nint} = \hat{U}(t) \prod_{E = 0}^{|e|V}
c^\dagger (E) | 0 \rangle = \hat{U}(t) \prod_{n= - \infty}^\infty
d^\dagger (nW) |0 \rangle, \label{NONINT}
\end{equation}
where $\hat{U}(t)$ is the time evolution operator, $|0 \rangle$ is
the Fermi sea in equilibrium, and $c^\dagger (E)$ creates an
electron with the single-particle energy $E$ in the scattering wave
$\psi(E)$ incoming from the source 1. $d^\dagger(X = nW)$ creates an
electron in the wave packet \cite{Martin,SINC}, $\phi(X) \propto
\int_0^{|e|V} dE \exp (-iEX/\hbar v_F) \psi(E)$, centered at
position $x=X$ with packet width $W = v_F \tau_V$. Here, $n$ is an
integer and we use the convention that both the arms amount to the
range $(-L/2,L/2)$. Thus for $-L/2<X<L/2$, the packet center is
located in the arm $u$ or $d$. The equivalence between the two forms
in Eq. (\ref{NONINT}) can be easily verified from the identity $\{
d(nW), d^\dagger(n'W) \} = \delta_{n,n'}$ and the Pauli exclusion.
While the former form is more commonly used, e.g., in the linear
response regime, the latter may provide more insightful
understanding of the ensemble of nonequilibrium electron density
configurations
and the resulting intrinsic dephasing mechanism, as shown below.
Later we use a combination of the two forms.

As time $t$ passes, $\phi(X)$ evolves to $\phi(X+v_F t)$. When a
packet arrives at the input splitter $a$, it propagates into the arm
$u$ with the probability $R_a$ or into the arm $d$ with $T_a$. This
process describes the shot noise \cite{Martin,Buttiker} and
generates a nonequilibrium ensemble of electron density
configurations in the arms. The ensemble depends on $N$, i.e., on
how many packets have significant weight in the arms. For example,
for $N \ll 1$, only one packet has significant weight and the
ensemble has two representative elements $m=1,2$, in each of which
the nonequilibrium density appears in the arm $u$ ($d$) with
probability $P_{m=1}=R_a$ ($P_{m=2}=T_a$) at any given time. In this
case, the nonequilibrium state has the form of $|\Psi \rangle_{\rm
nint} = \sum_{m=1,2} c_m |\Psi_m \rangle_{\rm nint}$ where $|c_m|^2
= P_m$ and $|\Psi_{m=1 (2)} \rangle_{\rm nint}$ contains the packet
state with significant weight in the arm $u$ ($d$) and no weight in
$d$ ($u$). Similarly, when two packets have considerable weight
[Figs.~\ref{E-MZIsetup}(b-e)], the ensemble has four representative
elements $m=1,2,3,4$, with probability $P_m = R_a^2$, $T_a^2$, $T_a
R_a$, and $R_a T_a$, respectively. When the interaction is turned
on, each density configuration causes different phase shift to an
interfering electron, resulting in the nonequilibrium dephasing.

{\it Lobe patterns in transmission probability.} --- We focus on a
weak-interaction regime where the energy relaxation rate is
sufficiently smaller than $t_{\rm fl}^{-1}$ \cite{Roulleau2}. In
this regime, a single-particle energy $E_0$ ($\in [0, |e|V]$) is a
well defined quantity, and thus one can consider the interference of
the scattering plane wave $|\psi(E_0) \rangle$. Within the arms,
$|\psi(E_0) \rangle$ shows the superposition, $|\psi(E_0) \rangle =
r|u(E_0) \rangle + t |d(E_0) \rangle$, of the plane waves $|l (E_0)
\rangle$ in the arms $l=u,d$, where $|r|^2 = R_a$ and $|t|^2 = T_a$.
The phase accumulation of $|l (E_0) \rangle$ is affected by the
interaction between $|\psi(E_0) \rangle$ and the rest of the
nonequilibrium electrons. We call the rest as the environment of the
electron in $|\psi(E_0) \rangle$.

To see how the interaction affects the phase accumulation, we write
the total nonequilibrium state as $|\Psi \rangle = |\Psi
\rangle_\textrm{nint} = |\psi(E_0) \rangle \otimes
|\Phi_{E_0}\rangle$. We will later turn on a weak interaction
between $\psi(E_0)$ and $\Phi_{E_0}$, and ignore the interaction
among the environment electrons, as a weak interaction may only
slightly modify the nonequilibrium densities from the noninteracting
case. Both $\psi(E_0)$ and the environment electrons in $\Phi_{E_0}$
are injected from the source $1$ and end in either the drain $3$ or
$4$. As $\Phi_{E_0}$ will be traced out to obtain the transmission
probability $\mathcal{T}(E_0)$ of $\psi(E_0)$ to a drain, one has
freedom of choosing either the wave-packet or the plane-wave
description for $\Phi_{E_0}$. We choose the former, since it is more
insightful and convenient for obtaining the density configurations
of $\Phi_{E_0}$ in the arms, which is essential to describe the
interaction. In the former, $\Phi_{E_0}$ is described by the moving
train of wave packets constructed by single-particle states with
energy $\in [0, |e|V]$ except $E_0$; the exclusion of $E_0$
negligibly modifies the packets from $\phi(X)$. As discussed before,
due to the shot noise at the splitter $j=a$, $|\Phi_{E_0} \rangle$
has the superposition, $|\Phi_{E_0} \rangle = \sum_m c_m
|\Phi_{E_0,m} \rangle$, of the multiple-packet states, each
corresponding to an element $m$ (with $P_m = |c_m|^2$) of the
nonequilibrium density ensemble. As the spatial configurations of
the moving packets are repeated in time with periodicity $\tau_V$,
time ensemble average over $[0, \tau_V]$, in addition to the average
over the nonequilibrium density ensemble, is necessary to obtain
$\mathcal{T}({E_0})$. We remark that the ensemble averages cannot be
captured by a mean-field approach.

We turn on a weak interaction between $\psi(E_0)$ and $\Phi_{E_0}$.
It causes the phase shift of $\Psi$ described by an operator,
$\hat{U}_{\rm ph} (t_0) = \sum_{l=u,d} e^{-i \frac{U_0}{\hbar}
\int_{t_0}^{t_0 + t_{\rm fl}} dt \hat{N}_l(t)} |l (E_0)\rangle
\langle l(E_0) |$. Here, $U_0$ is the interaction strength (assumed
\cite{INTERACTION} to be independent of $|e|V$), $t_0 \in [0,
\tau_V]$ is the time-ensemble index, and $\hat{N}_{l}(t) = \int_{x
\in l} \hat{n} (x,t)  dx$ acts on $|\Phi_{E_0} \rangle$ to measure
the number of the environment electrons in arm $l$ at time $t$. The
time ordering operator is dropped in $\hat{U}_{\rm ph}$ as the
packet dynamics $\phi(X+v_Ft)$ makes $[\hat{N}_l(t), \hat{N}_{l'}
(t')]$ a $c$-number \cite{Neder_TH}.

The interference signal of the interfering electron with
energy $E_0$ can be obtained from
the reduced density matrix
$\mathrm{Tr}_{\rm env} [ \hat{U}_{\rm
ph}(t_0) |\Psi \rangle \langle \Psi | \hat{U}_{\rm ph}^\dagger(t_0)]$,
which is found to be
\begin{eqnarray}
R_a |u (E_0) \rangle \langle u (E_0) | + T_a |d (E_0) \rangle \langle
d (E_0) | \nonumber \\
+ \sum_m P_m (r t^* \langle e^{i \hat{\delta} (t_0)}
\rangle_m |u (E_0) \rangle
\langle d (E_0) | + \mathrm{h.c.}),
\label{DENSITYM}
\end{eqnarray}
where $e^{i \hat{\delta}} \equiv e^{-i \frac{U_0}{\hbar} \int^{t_0 +
t_{\rm fl}}_{t_0} dt [\hat{N}_u(t) - \hat{N}_d(t)]}$ and $\langle
\cdots \rangle_m \equiv \langle \Phi_{E_0,m}| \cdots | \Phi_{E_0,m}
\rangle$. The trace $\mathrm{Tr}_{\rm env}$ over all the orthogonal
multiparticle states $\{ |\Phi_{{\rm env},i} \rangle \}$ of the
environment electrons is evaluated using the fact that
$|\Phi_{E_0,m}\rangle$ and $|\Phi_{E_0,m'\ne m} \rangle$ are
``severely'' orthogonal to each other, i.e., $\langle \Phi_{{\rm
env},i} | \prod_{k=1}^{k_{\rm mx}} \hat{n}(x_k) | \Phi_{E_0,m}
\rangle \langle \Phi_{E_0,m'} | \prod_{k'=1}^{k'_{\rm mx}}
\hat{n}(x_{k'}) | \Phi_{{\rm env},i} \rangle$ is nonzero only for $m
= m'$ where $k_{\rm mx}, k'_{\rm mx}$ are finite positive integers;
any finite number of local density operators can not transform a
packet of finite width into another orthogonal packet. The
off-diagonal part of the reduced density matrix in Eq.
(\ref{DENSITYM}) describes the dephasing of $|\psi(E_0)\rangle$ due
to the interaction.

The factor $\langle e^{i \hat{\delta}} \rangle_m$ can be further
evaluated as
\begin{eqnarray}
\langle e^{i \hat{\delta} (t_0) } \rangle_m \simeq
e^{i \langle \hat{\delta} (t_0) \rangle_m}
= e^{-i \frac{U_0}{\hbar}
\int_{t_0}^{t_0 + t_{\rm fl}} dt \langle \Delta \hat{N}(t) \rangle_m},
\label{PHASE}
\end{eqnarray}
where $\Delta \hat{N}(t) \equiv \hat{N}_u (t) - \hat{N}_d (t)$. In
Eq.~(\ref{PHASE}), we ignore the number fluctuation,
$\langle \Delta \hat{N}(t) \Delta \hat{N}(t') \rangle_m - \langle
\Delta \hat{N} (t) \rangle_m \langle \Delta \hat{N} (t') \rangle_m$,
or $\langle \hat{\delta}(t_0) \hat{\delta}(t_0) \rangle_m - \langle
\hat{\delta} (t_0) \rangle^2_m$, based on the observation that it is
not a crucial factor for the nonequilibrium dephasing \cite{FLUC}.


From Eq. (\ref{DENSITYM}),
the transmission probability of the electron in $\psi(E_0)$
to the drain $i=3$
is obtained as
\begin{eqnarray}
\mathcal{T} \equiv
\mathcal{T}(E_0)
= \mathcal{T}_0 + \mathcal{T}_1 D \cos (\Phi_B + \eta_D),
\label{TRANSMI}
\end{eqnarray}
where $\mathcal{T}_0 = T_a T_b + R_a R_b$,
$\mathcal{T}_1 = 2 \sqrt{T_a T_b R_a R_b}$, $\Phi_B$ is the
Aharonov-Bohm phase,
$D \cos (\Phi_B + \eta_D) \equiv \langle \Re [ e^{i \Phi_B}
\sum_m P_m \langle e^{i \hat{\delta}(t_0)} \rangle_m] \rangle_{t_0}$,
$\langle \cdots \rangle_{t_0}$ means the ensemble average
over time $t_0 \in [0, \tau_V]$, $D$ is the nonequilibrium
dephasing factor, and $\eta_D$ is the phase shift of
$\mathcal{T}$. Notice that $\mathcal{T}$ is independent of $E_0$
when $L_u = L_d$ and that Eq. (\ref{TRANSMI}) reproduces
the noninteracting result when $U_0 = 0$.

\begin{figure}[t]
\includegraphics[width=0.45\textwidth,height=0.18\textheight]{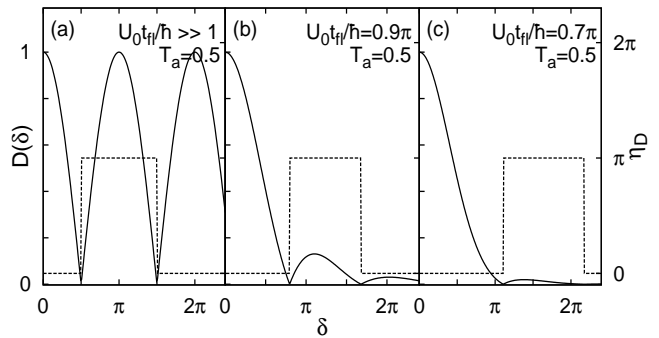}
\caption{Nonequilibrium dephasing factor $D(\delta)$ (solid curves)
and phase shift $\eta_D(\delta)$ of $\mathcal{T}$ (dashed). In (a),
we plot Eqs. (\ref{Lobe1},\ref{Phasejump1}), while (b) and (c) show
the numerical results; see text. In (b), $N$ varies from 0 to 2 as
$\delta$ increases up to $ 1.8 \pi$.
} \label{LobeInT}
\end{figure}

We first consider the regime of $N \ll 1$ and $U_0 t_{\rm fl}/\hbar
\gg 1$. Though this parameter regime is unphysical
\cite{INTERACTION}, it is nevertheless illustrative since it allows
the analytic evaluation of $D$ and $\eta_D$. In this regime, the
packet $\phi$ may be approximated as a square packet of extension
$W$ with constant density of $1/W$, and the nonequilibrium density
ensemble has two representative elements, one packet partially in
the $l=u$ arm with $P_{m=1} = R_a$ or in $d$ with $P_2=T_a$, as
discussed before.
Then, $\langle \hat{\delta}(t_0) \rangle_m$ in Eq. (\ref{PHASE})
has the constant value of $(-1)^m \delta$,
\begin{eqnarray}
\delta \equiv N \frac{U_0 t_{\rm fl}}{\hbar} = \frac{|e|V t_{\rm fl}}{2
\pi \hbar} \frac{ U_0 t_{\rm fl}}{\hbar}, \label{DELTA}
\end{eqnarray}
and one finds, using $\textrm{Arg}[\cdot] \equiv \arctan
(\textrm{Im} \cdot / \textrm{Re} \cdot)$, 
\begin{eqnarray}
D & = & \sqrt{\cos^2 \delta + (R_a - T_a)^2 \sin^2 \delta},
\label{Lobe1} \\
\eta_D & = & \textrm{Arg} [ \cos \delta - i (R_a - T_a) \sin \delta
]. \label{Phasejump1}
\end{eqnarray}
Notice that $\delta$ is proportional to the bias $V$ and that $D
(\delta)$ (thus $\mathcal{T}$) shows {\em lobe patterns} with
periodic minima of value $|R_a - T_a|$ at $\delta = \pi/2, 3 \pi/2,
\cdots$ [Fig.~\ref{LobeInT}(a)]. The minimum value increases from
zero as $T_a$ deviates from 0.5. And, the minima are accompanied by
{\em phase jumps}. At $T_a = 0.5$, $\eta_D$ shows sharp jumps of
$\pi$ at the minima, while it is zero or $\pi$ otherwise. The jumps
become smeared as $T_a$ deviates from 0.5. All these features are
related to the which-path information \cite{Feynmann} of the
nonequilibrium electrons. As $|e|V$ increases, the phase shift
$\langle \hat{\delta} \rangle$ increases (decreases) by $\delta$
with probability $T_a$ ($R_a$) as the environment electrons are in
arm $d$ ($u$). At $T_a = 0.5$, the two possible phase shifts in the
opposite direction are balanced in probability, resulting in the
sharp jumps in $\eta_D$ at the lobe minima. When $T_a \neq 0.5$, the
balance is broken, leading to the smearing of the jump.


Hereafter, we discuss numerical results for the weak interaction
regime with $U_0 t_{\rm fl}/\hbar < \pi$
\cite{INTERACTION}. A quantitative
calculation needs to take account of a larger number of
nonequilibrium ensemble elements than Fig.~\ref{E-MZIsetup} might
imply, since the packet $\phi$ has a rather slowly decaying tail and
more importantly for given $N$, about $2N$ packets appear in the
arms during $t_{\rm fl}$. Thus the minimum number of ensemble
elements for a given $N$ is $2^{2N}$; in the calculation,
$2^{9}$ elements are considered in the regime of $N \le 3$. Such
variety (of $\langle e^{i \hat{\delta} (t_0)} \rangle_m$) modifies
the lobes [Figs.~\ref{LobeInT}(b,c)]. Interestingly, the lobes in
$D$ and the jumps in $\eta_D$ of the $N \ll 1$ case are maintained,
though the lobes now acquire a decaying envelope and the minimum
positions of the lobes are shifted. This robustness can be
understood as follows. Among the $2^{2N}$ elements of the
nonequilibrium density ensemble, those with probability
$P_m=R_a^{2N}$ or $T_a^{2N}$ have $\langle \Delta \hat{N} \rangle_m
= N$ or $-N$, generating the same phase shift of $\pm \delta$ at all
$t_0$'s as in the $N \ll 1$ case, while in the others $\langle
\Delta \hat{N} \rangle_m$ varies with $t_0$ within a range smaller
than $\delta$. After the ensemble average over $t_0$, the lobes
themselves survive due to the former, while the latter has less
contribution to $\mathcal{T}$, giving rise to the decaying envelope
and the shift of the lobe positions. For smaller $U_0 t_{\rm
fl}/\hbar$, larger number of packets are involved (at a given
$\delta$), resulting in more rapidly decaying envelope and thus
smaller number of visible lobes [Fig.~\ref{LobeInT}(c)].

{\it Lobe patterns in dI/dV.} ---
From the zero-temperature current
$ I = (|e|/h) \int^{|e|V}_0 dE_0 \mathcal{T}(E_0) = (e^2/h) V
\mathcal{T}$,
one evaluates $dI/dV = (e^2/h) \mathcal{T}_0 [1 + F(V) \cos (\Phi_B
+ \eta_F(V))]$. Here, $F(V)$ gives the visibility of $dI/dV$,
$(dI/dV|_{\rm max} - dI/dV|_{\rm min})/(dI/dV|_{\rm max} +
dI/dV|_{\rm min})$, and $\eta_F$ is the phase shift of $dI/dV$;
remember $\delta \propto V$ and $L_u=L_d$.

\begin{figure}[t]
\includegraphics[width=0.45\textwidth,height=0.20\textheight]{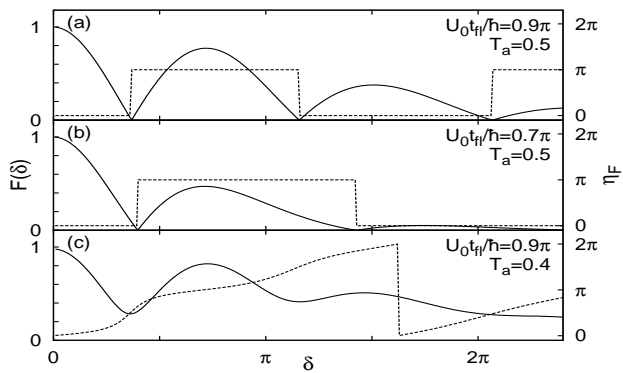}
\caption{Visibility $F$
(solid curves) and phase shift $\eta_F$ (dashed)
of $dI/dV$, as a function of $\delta$.
}
\label{LobeInDIDV}
\end{figure}

The lobes in $\mathcal{T}$ give rise to
similar lobes in $dI/dV$.
At $T_a = 0.5$, the visibility $F(V)$ shows a lobe
pattern whose minima
reach zero
and the phase $\eta_F$ jumps by $\pi$ at the minima of the lobes
while staying constant in other regions.
The number of visible lobes
in $F(V)$ becomes smaller for smaller $U_0 t_{\rm fl}/\hbar$,
similar to $D$
[Figs.~\ref{LobeInDIDV}(a,b)].
The first zero of $F(V)$ appears even for small
$U_0 t_{\rm fl}/\hbar$ where the second lobe
of $\mathcal{T}$ almost vanishes [Figs.~\ref{LobeInT}(c) and
\ref{LobeInDIDV}(b)].
As $T_a$ deviates from 0.5,
the minimum values of $F(V)$ increase from zero, similarly
to those of $D$, and the jumps of $\eta_F$
become smeared. In this case
the jump around the first minimum of $F(V)$
is sharper than that around the second [Fig.~\ref{LobeInDIDV}(c)].

{\it Discussion.} --- The lobes and phase jumps in
Fig.~\ref{LobeInDIDV}
qualitatively agree with the experimental
data \cite{Neder_EXP,Roulleau}
at $T_a = 0.5$, where $N
\simeq 0.4 - 1$ at $V = 8 \,\, \mu \textrm{V}$ in the E-MZI with $L
\simeq 10 \,\, \mu \textrm{m}$ and $v_F \simeq (2-5) 10^4 \,\,
\textrm{m/s}$.
The first minimum of $F(V)$ occurs around $\delta \simeq 3\pi/8$
(where $N \simeq 0.4$) in
Fig.~\ref{LobeInDIDV}, while around $V = 8 \,\, \mu $V in
the data. From this, we estimate
$U_0 \simeq 1.5 - 9 \,\, \mu e$V for the experimental E-MZIs.
The dependence of the number of visible lobes in $F(V)$ on
$U_0 t_{\rm fl}/\hbar$ \cite{INTERACTION} indicates
that the number depends
on the magnetic field, disorder, equilibrium electron density, etc,
as in the data.
Note that the minima of $F(V)$ look periodic in $V$
in Fig.~\ref{LobeInDIDV}(a),
although not in general.

The case with $\Delta L \equiv L_u - L_d \ne 0$ can be understood
from Eqs. (\ref{TRANSMI},\ref{Lobe1},\ref{Phasejump1}).
$D$ and $\eta_D$ have the same forms as in Eqs.
(\ref{Lobe1},\ref{Phasejump1}) except for the replacements, $\delta
\to \alpha V (L_u + L_d)/2$ and $\eta_D \to \eta_D + E_0 \Delta L /
\hbar v_F - \alpha V \Delta L /2$, where $\alpha = |e| U_0 t_{\rm
fl} / (2 \pi \hbar^2 v_F)$. Then, the shift of the lobe-minimum
positions is governed by $(L_u + L_d)/2$, and for $U_0 t_{\rm
fl}/\hbar < \pi$, $\eta_F$ becomes an undulating function that no
longer shows sharp jumps.
These modifications are however
negligible for $\Delta L \ll W = 2 \pi \hbar v_F / (|e|V)$, in
agreement with experimental data \cite{Neder_EXP}.


We suggest that the combined experimental analysis of both $I$ and
$dI/dV$ may be useful since $I$ can provide the direct information
of the dephasing factor $D$. And, our formalism may be applicable to
possible nonequilibrium dephasing in other electronic
interferometries \cite{Neder_EXP2}. Finally, careful treatment of
the number fluctuation \cite{FLUC}, ignored in Eq.~(\ref{PHASE}),
may give further understanding, as it may modify the decaying
envelope and the lobe positions.


In summary, we have shown that the nonequilibrium density ensemble,
generated by the shot noise at the input beam splitter, can
cause the nonequilibrium dephasing in the E-MZI.
Our result suggests that
the experimental data in Ref.~\cite{Neder_EXP} may
only be the first of its kind with more nonequilibrium
quantum effects waiting for their discoveries.

We thank I. Neder for sending us experimental data
and M. B\"{u}ttiker for useful discussions. We are supported by
KRF (2006-331-C00118, 2005-070-C00055) and by MOST through the leading basic S \&
T research projects.

{\it Note added.}--- During our manuscript preparation,
a preprint \cite{Neder_TH2} addressing shot-noise effects in the
E-MZI was reported. It considers a regime of the interaction
strength stronger than ours.

\bibliographystyle{apsrev}

\begin{thebibliography}{99}
\bibitem{INTERACTION}
One may have $U_0 \sim e^2 / (\epsilon L)$, where $\epsilon$ is the
dielectric constant of the arms. Then, $U_0 t_{\rm fl} / \hbar \sim
e^2 / (\epsilon \hbar v_F)$ is independent of $|e|V$ and $L$,
justifying our assumption of $U_0$ being independent of $V$. And, we
choose $U_0 t_{\rm fl} / \hbar < \pi$, where single-particle level
spacing $\hbar \pi / t_{\rm fl}$ is larger than $U_0$ so that the
E-MZI is in the weak interaction regime.
\bibitem{FLUC}
When $\hat{N}_l(t)$ is projected onto the energy window $[0,|e|V]$,
the number fluctuation is indeed negligible. Even when the virtual
transition to outside this energy window caused by $\hat{N}_l(t)$ is
taken into account, it may not affect the lobe pattern qualitatively
since the $m$-dependence of this fluctuation is negligible, though
it may modify the decaying envelope and the lobe minimum positions.
\end{thebibliography}

\end{document}